\documentclass[12pt]{article}
\usepackage[utf8]{inputenc}

\begin{document}

\title{Using Recurrent Neural Networks to Predict Aspects
of 3-D Structure of Folded Copolymer Sequences}

\author{Ronan G. Reilly$^1$, M-Tahar Kechadi$^1$,\\
Yuri A. Kuznetosov$^2$, Edward G.
Timoshenko$^2$\thanks{E-mail: Edward.Timoshenko@ucd.ie}, \\
and Kenneth A. Dawson$^2$
}

\date{{\small
Department of Computer Science$^1$\\
and Department of Chemistry$^2$,\\
University College Dublin, Ireland
}}

\maketitle

\begin{abstract}
The neural network techniques are developed for artificial sequences based on approximate models of proteins. We only encode the hydrophobicity of the amino acid side chains without  attempting to model the secondary structure. We use our approach to obtain a large set of sequences with known 3-D structures for training the neural network. By employing recurrent neural networks we describe a way to augment a neural network to deal with sequences of realistic length and long-distant interactions between the sequence regions. 
\end{abstract}


\section*{Introduction}

Neural networks have been applied with some limited success to the problem
of predicting the secondary [1,2] and tertiary [3,4] structure of proteins based on their
amino acid residue sequence. The number of sequences for which there is a known 3-D structure is relatively limited. The rate at which 3-D structures are being solved is
at least one order of magnitude lower than the rate at which new protein sequences
are being determined [3]. In addition, a limitation in the neural network approaches
taken to date is their inability to deal with very long sequences, and with the
possibility of dependencies between different regions of a sequence [8]. The work
described here is an attempt to address these limitations. In order to obtain a large set
of sequences with known 3-D structures for training the neural network, we use the
approach described in [5] to generate a set of artificial copolymers consisting of
hydrophobic and hydrophilic units with a known 3-D structure when folded. By
employing recurrent neural networks and building on the approach described in [3, 4],
we describe a way to augment a neural network with both with a facility to deal with
sequences of realistic length, and with a mechanism for handling possible long-distant interactions between regions of the sequence.
These sequences are very approximate models of real proteins, given that we
only encode the hydrophobicity of the amino acid side chains, and there is no attempt
to model their secondary or super-secondary structure. Nonetheless, the neural
network techniques developed using artificial sequences are readily applicable to real
proteins.

\section*{Sequence generation}

In order to study the structure of copolymer folding we use the painting
method described in [5]. The general approach is as follows:
\begin{quotation}
$ \quad $\\
1. An open homopolymer sequence of a fixed size is chosen;\\
2. This homopolymer sequence is allowed to fold into a globule using a
Monte Carlo simulation method [8];
SRNs and Protein Folding\\
3. The globule shape depends on the position of each monomer in the
sequence and the interactions between them. Since this globule is not
always compact, only spherical globules are considered for the next stage;\\
4. The core of the globule is painted. The volume of the coloured core is
determined by the desired hydrphobicity ratio along the chain. A typical
ration is 50:50. The technique used to do the painting is called the regular
hull painting algorithm;\\
5. The coloured sequences obtained are considered as copolymer sequences.
The Monte Carlo simulation method is then used to refold the copolymer
sequences to determine how compactly they collapse.
\end{quotation}

In the case of this experiment, a subset of 500 of the sequences generated by
this painting technique were coded as hydrophobicity and repeatedly refolded using
the Monte Carlo simulation technique. The quality of the refolded sequence was
measured by the disorder parameter $\Psi$:

\[
\Psi = \frac{1}{2N^2} \sum_{ij} (\lambda_i + \lambda_j)D_{ij}
\]
\[
\mbox{where } D_{ij} = \frac{1}{3}\left((x_i - x_j)^2 + (y_i - y_j)^2 + (z_i - z_j)^2\right),
\]
\[
\mbox{and } \lambda = \biggl\{
\begin{array}{ll}
-1 \mbox{\ for hydrophobic} \\
+1 \mbox{\ for hydrophillic}
\end{array}
\biggr.
\]

For an ideally folded sequence with hydrophobic units on the inside of the globule
and hydrophilic on the outside, the value for $\Psi$ tends to zero. Any deviation from this
ideal gives larger positive or negative values.
As a result of the refolding experiments it was possible to assign a given
sequence to one of three categories: sequences that folded well (60\%), sequences that
folded poorly (37\%), and sequences that folded very poorly (3\%). In the case of the
poor folders, these often comprised a globule with a trailing strand, or in some cases
two small globules. Thus, they appeared to be moving in the right direction, and
given more iterations in the Monte Carlo simulation would eventually produce a
compactly folded globule.

\section*{SRN architecture}

A simple recurrent network (SRN) is an extension to the standard feedforward neural
network and was first proposed by Elman [6] (cf. Figure 1). Its main use to date has
been in the area of modelling time series data and the acquisition of grammar. A
common feature of both sentence structures and protein structures is the inherent
sequential nature of the ``input”, and the sometimes non-sequential nature of the inter-relationships between different regions of the sequence. So, for example, in a
sentence ``The cat who chased the dog bites the mouse” there is agreement in number
between the first noun ``cat” and the verb ``bite”. This is referred to as a long-distance
dependency. Similar long-distance interactions are also possible between the sub-structures of a folded protein [8]. Furthermore, a significant strength of SRNs is their
ability to deal with sequences of varying and potentially unlimited length.


The main difference between the SRN architecture and a regular feed-forward
neural network is that it permits the state of the hidden units at the previous time step
to be part of the input at the next time step. This provides the network with an
attenuated memory of its inputs at preceding time steps (and not just the most recent
one), and permits it to use information from earlier inputs in the processing of current
ones.

\section*{A General Input Representation}

As well as choosing the correct architecture for carrying out protein structure
prediction, a key factor in the success or otherwise of any artificial neural network
approach to this challenge is the way in which the problem is presented to the
network, particularly the choice of input representation. In addition to exploring the
use of SRNs, another goal of this project is to try to develop an input representation
that provides as much information as possible within the constraints of the number of
input units available.


In both of the application to be discussed below, the input to the network is in
the form of a moving window through which a given subset of the copolymer
sequence is input. Such an input technique is similar to that employed by [1] and [3].
However, given that we will be restricting ourselves to binary representations of
polymer sequences where a hydrophilic unit is coded as ``1” and a hydrophobic unit is
coded as a ``0”, there is the possibility of using a moving window with a number of
different resolutions (cf. Figure 2). The number of input units used in the two
applications described here is 50 for the elements of the sequence, plus one used to
indicate when the entire sequence has been input to the network. This latter unit is
referred to as the reset unit, because when it is switched on, the feedback from the
hidden units is reset to a vector of zeros.
For the central 10 units of the input , there is a one-to-one mapping between
sequence elements and the input units of the network. However, as one moves away
from the centre, to the right and left, the resolution of the input units is reduced. Thus
in the adjacent blocks of the 10 units, each input unit codes the average of four binary
units. In the next pair of blocks, the units code for the average of 10 units. This gives
a form of ``fisheye” lens effect which permits an effective window size of 280, albeit
with diminishing resolution as one moves away from its centre.

\section*{The Experiments}

The purpose of the following two experiments was to test the feasibility of
using an SRN to predict various aspects of the structure of a folded copolymer. The
first experiment was designed to predict the propensity of the sequence to refold
successfully, that is to which of the two refolding classes the sequence belonged.
The second experiment attempted the more ambitious goal of predicting the 3-D structure of those sequences that refolded the best (i.e., with a value for $\Psi$ of near
zero). The structure of the folded sequence was represented to the network in the
form of a binary distance inequality matrix (DIM). Both experiments can be
considered as preliminary steps towards developing a system to predict the 3-D
structure of real proteins.

\subsection*{Materials}

The initial set of materials used in both experiments comprised 500 painted
and refolded copolymers of length 120, produced using the painting technique
described earlier. All sequences consisted of a series of 1s and 0s, with ``1”
representing a hydrophilic element and ``0”, a hydrophobic one.

\subsection*{Refolding ability}

In the first set of experiments, a simple recurrent network was used to predict
the refolding class of the open co-polymer. Only two classes were used: good and
poor. Because of the low percentage of very poor folding sequences, it was not
feasible to train the network on this class of data. The input sequence was pre-
processed through a multi-resolution window as described in the previous section.
The task of the SRN was to switch on one of two output units indicating to which
folding class the sequence belonged.

The prediction network consisted of 51 input units, 10 hidden units\footnote{A number of hidden unit values were tested (80, 40, 20), and 10 proved to be optimal.}, 10
feedback units, and 2 output units. This network was trained on 100 folded and
painted sequences using the backpropagation learning algorithm [7], with the learning
rate parameter set to 0.01, and the momentum parameter to 0.5. The training patterns
for the SRN consisted of a series of vectors generated from the multi-resolution
window as it was shifted across the open sequence, 10 units at a time. This meant
that a sequence of 120 would take 12 input presentations to be completely presented
to the network. The activation values from the network’s hidden units at the previous
time step was provided as input, in addition to each window vector. When the
window came to the end of a sequence, the hidden unit feedback was set to a vector
of zeros. For each of the 12 window presentations, the target for the two output units
was the same: the refolding class of the input sequence. The network was trained for
300 complete passes (epochs) through the training set.

  \begin{table}
\centering
\begin{tabular}{|c|c|c|}
\hline
& \multicolumn{2}{c|}{{\bf Refolding class}} \\
\cline{2-3}
{\bf Network prediction} & {\bf Good } & {\bf Bad} \\
\hline
{\bf good} & p(g$\vert$G) = 0.73 & p(g$\vert$B) = 0.27 \\
\hline
{\bf bad } & p(b$\vert$G) = 0.59 & p(b$\vert$B) = 0.41 \\
\hline
\end{tabular}
\caption{
Performance of the SRN in predicting the quality of refolded sequences it has not been trained on. The values in cells represent conditional probabilities, where ``g'' is the network prediction and ``G'' is the actual classification of the sequence. The network was trained on 100 sequences, and then tested with an additional 100.
}
\label{tab:accuracy}
\end{table}


\subsection*{Refolding prediction results}

The real valued output of this network was converted to a single binary
decision value by assuming the largest of the two outputs from the network to
indicate a classification decision (i.e., good or bad refolder). This was then compared
to the actual classification of the sequence. The results in Table 1 are representative
of the results of a number of training replications and are given in the form of
conditional probabilities. Typically, after 300 or so passes through the training set,
the network learned to perform the classification task with 100\% accuracy. It was
then tested on an additional 100 sequences not seen during training. As can be seen,
while the network is good at predicting good folding sequences, it tends to
misclassify badly folding sequences as good ones.


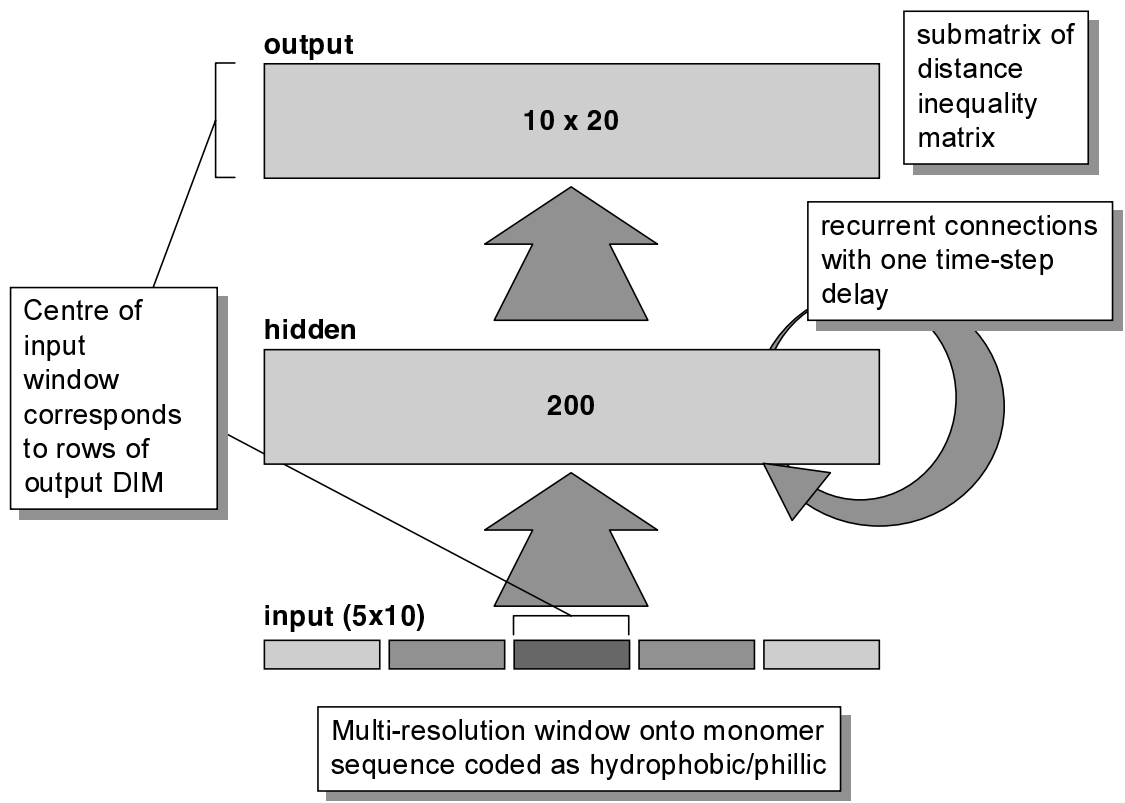
\subsection*{Distance inequality matrix prediction}

A distance inequality matrix (DIM) is a binary version of the 2-D matrix of
Euclidean distances between each element and every other in the sequence. In this
study, distances greater than 3 monomers were coded as 0.5, less than 3 as -0.5. The
task of the network was to output a series of 10x20 sub-matrices of the overall
distance matrix for a given input sequence in row major order. The overall
architecture of the SRN is given in Figure 3.


As with the earlier refolding prediction experiment, the SRN was trained by
inputting a binary coded protein using a moving window across the entire sequence.
The structure of the input window has already been described (see Figure 2). At its
centre is a 10 unit one-to-one representation of the sequence fragment, while the
elements either side of the window are represented with diminishing resolution.
However, in the case of the DIM prediction task, each window was used to generate a
sequence of 10x20 sub-matrices of the overall DIM (see Figure 4). These sub-matrices 
corresponded to the rows of the DIM associated with the 10 elements of the
sequence at the centre of the window. Rather than getting the network to output an
entire row, it was generated in a series of steps. The 10x20 sub-matrices partially
overlap (by 50\%), thus providing some element of redundancy in the construction of
the overall matrix. The matrix is also symmetric about the diagonal, and this was also
exploited in constructing a full DIM for a given sequence.
From a set of 200 refolded sequences a subset of the best refolders was
selected. These were then divided into a training set and a test set of (70 and 69
respectively). An SRN comprising 51 input units, 80 hidden units, and 200 output
units was trained for 500 epochs to predict the DIM for the 100 sequences in the
training set. A learning rate of 0.01 and a momentum of 0.5 was used. The network
was then tested on its ability to predict accurately the DIM for the set of 100
sequences that it had not been trained on.


\subsection*{Distance matrix results}

Testing the DIM prediction network involved reconstructing a matrix from the
real-valued network output, converting it to binary form and then comparing it to the
actual DIM for that sequence. As has already been mentioned, both the diagonality of
the matrix and the 50\% overlap between the adjacent 10x20 output sub-matrices was
exploited in this reconstruction. In both the diagonal and overlap cases, where there
were multiple estimates for a given cell of the DIM, an average of these values was
used. The averaging of the overlap was done prior to averaging across the diagonal.
After averaging, the reconstructed matrix was converted to a DIM by thresholding the
values at 0.5. The percentage of elements of the target and output DIMs that are in
the same state (i.e., on or off) is then calculated and used to measure prediction
accuracy.


\begin{table}
\centering
\begin{tabular}{|c|c|}
\hline
\multicolumn{2}{|c|}{{\bf Accuracy of prediction}} \\
\hline
{\bf training set (n=70)} & {\bf test set (n=69) }\\
\hline
80\% & 77\% \\
\hline
\end{tabular}
\caption{
Performance of SRN in predicting the distance inequality matrix of refolded copolymer sequences of length 120. The figures represent the average prediction performance over the n sequences comprising each set.
}
\label{tab:refolding_prediction}
\end{table}

As can be seen from Table 2, the training performance achieves an 80\% level
of accuracy, while a fairly respectable 77\% accuracy is attained on the unseen test set
of sequences. In evaluating these results it should be borne in mind that the there are
a number of equally acceptable folded configurations for any of the good sequences
used in this task. Therefore it is not possible to predict precisely what configuration a
sequence will adopt when folded. The best way to view what this network does is
predict a configuration that is typical for a particular type of sequence.

\section*{Discussion}

The results of these experiments have shown that while SRNs have a poor
ability to determine the refoldability of a given copolymer sequence, they appear to
be able to predict with some degree of accuracy the 3-D structure of folded
copolymers. In both cases, predictions are based on the hydrophobicity of sequence
elements. In the first experiment, the network proved more reliable in its prediction
of good folders than poor ones. This suggests that determining the badness of a
folding is based on fairly subtle and relatively inaccessible information in the
sequence. On the other hand, the second experiment suggests that the use of the DIM
prediction framework seems a viable technique for accurately predicting a structure
for folded copolymers. Moreover, the technique presented are not limited to any
particular sequence length, and could be trained on a sequence of different lengths.

\section*{Future work}

The next goal of the research described here, is to generalise the results to
longer sequences, and to apply the techniques described here to real protein
sequences with secondary as well as tertiary structure. Bohr and his co-workers [3,4]
have used non-recurrent feed-forward networks which takes a sub-sequence of the
amino acid residue sequence in an input 
window\footnote{Bohr and colleagues use a window with uniform resolution, unlike the method employed here.}
and outputs both the rows of the
DIM associated with the residue in the window’s centre and the class of secondary
structure (e.g., $\alpha$-helix, $\beta$-sheet) to which it belongs. On the basis of the research
described here, it is conjectured that it may not be necessary to explicitly train the
SRN to recognise the secondary structure of the subsequence, given that the window
provides a relatively large view of the sequence neighbourhood compared to Bohr’s
method, and thus may permit the implicit ``recognition” of secondary structure.


\begin{thebibliography}{9}
\bibitem{1}
Qian, N., and Sejnowski, T.J. (1988). Predicting the secondary structure of
globular proteins using neural network models. J. Mol. Biol., 202, 865-884.
\bibitem{2}
McGregor, M. J., Flores, T. P., \& Sternberg, M.J.E. (1989). Prediction of beta
turns in proteins using neural networks. Protein Eng., 2, 521-526.
\bibitem{3}
Bohr, J., Bohr, H., Brunak, S., Cotterill, R.M.J., Fredholm, H., Lautrup, B.,
Petersen, S.B. (1990). A novel approach to prediction of the 3-dimensional
structures of protein backbones by neural networks. FEBS Letter, 261, 43-46.
\bibitem{4}
Bohr, J., Bohr, H., Brunak, S., Cotterill, R.M.J., Fredholm, H., Lautrup, B. and
Petersen, S.B. (1993) Protein Structures from Distance Inequalities. J. Mol. Biol.,
231, 861-869.
\bibitem{5}
Kechadi, M-T, Reilly, R.G., Kuznetsov, Y., Timoshenko, E., Dawson, K.  
A study of sequence distributions of a painted globule as a model for
proteins with good folding properties. {arXiv:cond-mat/0104440.} 
\bibitem{6}
Elman, J.E. (1990). Finding structure in time. 
Cognitive Science, 14, 179-211.
\bibitem{7}
Rumelhart, D. E., Hinton, G. E., \& Williams, R. J. (1986). Learning internal
representations by error propagation. In D. E. Rumelhart, J. L. McClelland, \&
The PDP Research Group (Eds.), Parallel distributed processing. Explorations in
the microstructure of cognition. Volume 1: Foundations. Cambridge, MA: MIT
Press.
\bibitem{8}
Klinger, T.M., \& Brutlag, D.L. (1994). Discovering structural correlations in
alpha helices. Protein Science, 3, 1847-1857.
\bibitem{9}
Binder, K. (1987). Applications of Monte Carlo methods in statistical physics,
2nd Edition. Berlin: Springer-Verlag.
\end{thebibliography}
\end{document}